\documentclass[11pt,a4paper]{article}
\pdfoutput=1
\usepackage[latin1]{inputenc}
\usepackage{jcappub}
\usepackage{amsmath}
\usepackage{amsfonts}
\usepackage{amssymb}
\usepackage{amsthm}
\usepackage{graphicx}
\usepackage{epsfig}
\usepackage{subfigure}
\usepackage{color}
\usepackage{amssymb}
\usepackage{bbold}
\usepackage{dsfont}
\newcommand*\diff{\mathop{}\!\mathrm{d}}

\def\beq{\begin{equation}}
\def\eeq{\end{equation}}
\def\baq{\begin{eqnarray}}
\def\eaq{\end{eqnarray}}

\title{CMB spectral distortions in generic two-field models}

\author[a,c]{Kimmo Kainulainen,}
\author[a,c]{Juuso Leskinen,}
\author[a,c]{Sami Nurmi,}
\author[b]{and Tomo Takahashi}

\affiliation[a]{Department of Physics, University of Jyv\"{a}skyl\"{a},  P.O. Box 35, FI-40014 University of Jyv\"{a}skyl\"{a}, Finland}
\affiliation[b]{Department of Physics, Saga University, 1 Honjo, Saga 840-8502, Japan}
\affiliation[c]{Helsinki Institute of Physics and Department of Physics, University of Helsinki, P. O. Box 64, FI-00014, Finland}

\abstract{We investigate the CMB $\mu$ distortion in models where two uncorrelated sources contribute to primordial perturbations.  We parameterise each source by an amplitude, tilt, running and running of the running. We perform a detailed analysis of the distribution signal as function of the model parameters, highlighting the differences compared to single-source models. As a specific example, we also investigate the mixed inflaton-curvaton scenario. We find that the $\mu$ distortion could efficiently break degeneracies of curvaton parameters especially when combined with future sensitivity of probing the tensor-to-scalar ratio $r$. For example, assuming bounds $\mu < 0.5 \times 10^{-8}$ and $r<0.01$, the curvaton contribution should either vanish or the curvaton should dominate primordial perturbations and its slow-roll parameter $\eta_{\chi}$ is constrained to the interval $-0.007 < \eta_{\chi}< 0.045$.}

\emailAdd{kimmo.kainulainen@jyu.fi}
\emailAdd{juuso.o.leskinen@student.jyu.fi}
\emailAdd{sami.t.nurmi@jyu.fi}
\emailAdd{tomot@cc.saga-u.ac.jp}

\begin{document}

\maketitle

%
\section{Introduction}
%

The large scale structure in the Universe originates from primordial fluctuations, which are thought 
to be generated from quantum fluctuations during inflation. In the minimal setup, they originate 
from a single scalar field, the inflaton, which drives the inflationary expansion. While the single 
field model is the leading candidate, it is quite possible that multiple fields were sourcing the 
perturbations. Additional scalars may participate in driving inflation or they may be 
``spectator" fields with no dynamical importance during inflation. However, quantum fluctuations 
acquired by light spectators can also source primordial perturbations. Examples of such 
models are the curvaton scenario~\cite{Moroi:2001ct,Enqvist:2001zp,Lyth:2001nq} and the modulated 
reheating~\cite{Dvali:2003em,Kofman:2003nx}.

We can probe fields responsible for primordial perturbations by studying their statistics via 
precision cosmological observations. In particular Planck~\cite{Ade:2015lrj} observations 
of cosmic microwave background (CMB), precisely determine the characteristics the primordial power 
spectrum. In addition, bounds on the tensor-to-scalar ratio $r$ and on non-Gaussianities are 
providing important constraints for inflationary models. Nevertheless, a wealth of inflationary 
models is still consistent with all current observations and more information is needed to 
remove degeneracies and pin down the correct model.
An obvious way to obtain more information is to study the detailed scale-dependence of primordial 
power spectrum. This is usually characterized by the scale dependence of the spectral tilt $n_s$, 
including the so called running $\alpha_s$, and the running of running $\beta_s$. Here information 
from a broad range of scales is needed, extending in particular to observations on the smallest 
scales, such as measurements of the 21cm 
fluctuations~\cite{Kohri:2013mxa,Munoz:2016owz,Sekiguchi:2017cdy} 
and of CMB spectral distortions~\cite{husilk:1994,Chluba:2012gq,Chluba:2012we,Dent:2012ne,Enqvist:2015njy,Cabass:2016giw,Cabass:2016ldu}. 

In this work we study $\mu$-type spectral distortions produced first, in a general two field model 
and second, in a specific mixed curvaton inflaton scenario. In the general setup we parametrize each 
field ($i=1,2$) with their own tilt and runnings, $n_i$, $\alpha_i$ and $\beta_i$. We scan the 
parameters with broad prior ranges and compute the $\mu$ distortion signal for models consistent 
with Planck	observations. We find that models with large differences $n_2-n_1$ and/or 
$\alpha_2-\alpha_1$ and an enhanced $\mu$ distortion can be accommodated by the current data.
In the context of the mixed inflaton-curvaton model we find that $\mu$ distortion combined 
with the upper bound on the tensor-to-scalar ratio $r$ can break degeneracies among model
parameters. For example finding $\mu < 0.5 \times 10^{-8}$ and $r<0.01$ would imply that most of the 
observed perturbations were sourced by the curvaton, with the curvaton slow-roll parameter constrained to range $-0.007 < \eta_{\chi}< 0.045$.

This paper is organized as follows. In section~\ref{sec:mudistortion} we briefly review the 
calculation of the $\mu$ distortion arising from photon diffusion. In section~\ref{sec:curvaturepert} we introduce the parametrization for the 
general 2-field model.  In section~\ref{sec:singletwosignal}, we present the 
strategy of our analysis and compute the $\mu$ distortion predictions.  In 
section~\ref{sec:modelcurv}, we study the specific example of the mixed inflaton and 
curvaton model. Finally section~\ref{sec:conclusion} contains our conclusions and outlook.

%
\section{Spectral distortions} 
\label{sec:mudistortion}
%

The spectrum of CMB photons cannot be completely thermal. It inevitably features small distortions generated below the redshift $z\simeq 2\times 10^6$, where interactions between electrons and photons become too slow to maintain a local equilibrium~\cite{Sunyaev:1970er, husilk:1994, Hu:1992dc}. Spectral distortions of $\mu$-type are generated for $z\gtrsim 2 \times 10^5$, when kinetic equilibrium is efficiently maintained by Compton scatterings but photon number changing processes through double Compton scattering and Bremsstrahlung are slow compared to the expansion rate, see~\cite{Chluba:2011hw,Sunyaev:2013aoa,Tashiro:2014pga} for recent reviews. Energy injection to photons through Silk damping of the acoustic waves leads to an excess of photons compared to chemical equilibrium. Below $z\simeq 5 \times 10^4$ also the kinetic equilibrium is lost and the distortions generated in this epoch are of the $y$-type~\cite{Chluba:2011hw}. The distortions of a more general type have been discussed in~\cite{Chluba:2011hw}.
In the intermediate epoch  $2\times 10^4 \lesssim z \lesssim 5 \times 10^5$ the kinetic equilibrium is partially maintained and a mixture of $\mu$ and $y$ distortions, the so called $i$ distortion, is formed~\cite{Khatri:2012tw}.

Full analysis of the distortions requires solving the Boltzmann equations for primordial plasma numerically~\cite{Chluba:2011hw}. Here we concentrate only on $\mu$ distortions whose time evolution can be approximated by~\cite{husilk:1994} 
\begin{equation}
\frac{\diff \mu}{\diff t} = - \frac{\mu}{t_{\mathrm{dC}}(z)} + \frac{1.4}{\rho_{\gamma}} \frac{\diff Q}{\diff t}\,.
\label{eq:mu-diff-evolution}
\end{equation}
Here \(\diff Q/ \diff t\) describes an energy injection to photons and \(t_{\mathrm{dC}}\) is the time-scale for the double Compton scattering. This assumes the dominant number changing processes are double Compton scatterings and that kinetic equilibrium is maintained through Compton scatterings. The approximative solution to eq.~\eqref{eq:mu-diff-evolution} is~\cite{husilk:1994} 
\begin{equation}
\mu = 1.4 \int^{z_2}_{z_1} {\diff z} \mathrm{e}^{-(z/z_{\mathrm{dC}})^{5/2}} \left(\frac{1}{\rho_{\gamma}} \frac{\diff Q}{\diff z}\right),
\label{eq:mu_integral}
\end{equation}
where \(z_{\mathrm{dC}} \approx 4.1 \times 10^{5} \left(1 - Y_{p}/2\right)^{-2/5} \left(\Omega_{b} h^{2}\right)^{-2/5}\)~\cite{husilk:1994} and \(Y_{p} \simeq 0.251\)~\cite{Ade:2015xua} denotes the primordial helium abundance. The lower limit of integration is $z_1 = 2 \times 10^6$. We neglect all intermediate distortions and assume instant transition between the $\mu$- and $y$-epochs, which sets the integral upper limit at  $z_2 = 5 \times 10^4$~\cite{Hu:1992dc}.

For distortions generated by the dissipation of acoustic waves the energy-injection rate in (\ref{eq:mu_integral}) is related to the spectrum of the primordial curvature perturbation $P_{\zeta}$ by~\cite{husilk:1994,Khatri:2011aj,Chluba:2012gq,Dent:2012ne}\footnote{See~\cite{Chluba:2012gq} for a discussion of the factor $3/4$.}:
\begin{equation}
\frac{1}{\rho_{\gamma}} \frac{\diff Q}{\diff z} = 
- \frac{3}{4} \int \frac{{\diff}^{3} k}{(2 \pi)^{3}} A_{\nu} P_{\zeta}(k) \frac{\diff \Delta^{2}_{Q}}{\diff z},
\label{eq:Einjection}
\end{equation}
where \(A_{\nu} \equiv 4/(2R_{\nu}/5 + 3/2)^{2}\) and \(R_{\nu} \equiv \rho_{\nu}/(\rho_{\gamma} + \rho_{\nu})\), with $\rho_{\nu}$ and $\rho_{\gamma}$ being the neutrino and photon energy densities and the differential energy injection term is
\begin{equation}
\frac{\diff \Delta^{2}_{Q}}{\diff z} =  \frac{\diff}{\diff z}\left( \frac{9 c^{2}_{s}}{2} \ \mathrm{e}^{-2 k^2/k^2_{\mathrm{D}}}\right) = 27 c^{2}_{s} A_{\mathrm{D}} \left(1 + z\right)^{-4} k^2 \mathrm{e}^{-2 k^2/k^2_{\mathrm{D}}}.
\label{eq:d_DeltaQ}
\end{equation}
Deep in the radiation dominated epoch and for constant sound speed, the photon diffusion damping scale $k_{\rm D}$ and the constant $A_{\rm D}$ above are given by~\cite{Khatri:2011aj}
\begin{align}
k_{\mathrm{D}} =  
\left(\frac{8}{135 H_{0} \Omega^{1/2}_{r} n_{e0} \sigma_{T}}\right)^{-1/2} \left(1 + z\right)^{3/2} \equiv A^{-1/2}_{\mathrm{D}} \left(1 + z\right)^{3/2} , \label{eq:diffscale_AdkD}
\end{align}
where $\sigma_{T}$ is the cross section of Thomson scattering, $H_0$ and $\Omega_{r}$ denote the current Hubble rate and radiation density and the free electron number density before the recombination is \(n_{e0} = n_{\mathrm{H}0} + 2 n_{\mathrm{He}0}\). 

The diffusion damping scales $k_{\rm D}(z_{1,2})$ corresponding to redshift integration limits in \eqref{eq:mu_integral} set the scales probed by the $\mu$ distortion: $ 50~{\rm Mpc}^{-1} < k < 10^4~{\rm Mpc}^{-1}$. These are much smaller than the scales probed by 
CMB anisotropies: ${\cal O}(10^{-3}) \ \mathrm{Mpc}^{-1} \lesssim k \lesssim {\cal O}(0.1) \ \mathrm{Mpc}^{-1}$ or large scale structure: ${\cal O}(10^{-2}) \ \mathrm{Mpc}^{-1} \lesssim k \lesssim {\cal O}(1) \ \mathrm{Mpc}^{-1}$, which makes the spectral distortion a very interesting observable.  While primordial black holes (PBH) also constrain similar small scale perturbations, their constraints on the spectrum are much less stringent~\cite{Josan:2009qn}. 

Finally, we note that in addition to the Silk damping part (\ref{eq:mu_integral}), there is an adiabatic cooling contribution from photon energy lost into heating up electrons after their decoupling, which yields a small negative part $\mu_{\rm ad} \approx -3 \times 10^{-9}$ to the total distortion~\cite{Chluba:2011hw,Khatri:2011aj}.

%
\section{General parameterisation of two-field models} \label{sec:curvaturepert}
%

In~\cite{Dent:2012ne,Cabass:2016giw} (see also \cite{clesse:2014}) it was found that in the case of single field inflation, future measurements of the $\mu$ distortion would place a powerful constraint on the running of the spectral index. Spectral distortions in mixed inflaton-curvaton models were further investigated in~\cite{Enqvist:2015njy}. Here we study generic two-field models and systematically investigate how future data on $\mu$ distortions constrains their parameter space. 

We consider a general two-field setup where the spectrum of primordial curvature perturbation, $\langle \zeta(\mathbf{k}) \zeta(\mathbf{k}^{\prime}) \rangle = (2 \pi)^{3} \delta(\mathbf{k} + \mathbf{k}^{\prime}) 2 \pi^2 \mathcal{P}_{\zeta}/k^3$ can be parameterised as
\begin{eqnarray}
\mathcal{P}_{\zeta} &=&  {\cal P}_1 + {\cal P}_2=
\frac{A_{s}}{1 + R} \left[\left(\frac{k}{k_{\mathrm{ref}}}\right)^{n_{1} - 1 
+ \frac{1}{2} \alpha_{1} \mathrm{ln}\left( \frac{k}{k_{\mathrm{ref}}} \right)
+ \frac{1}{6} \beta_{1} \mathrm{ln}^2 \left( \frac{k}{k_{\mathrm{ref}}} \right)}
\right.
\nonumber \\
&&\phantom{helluriijahelli} \left.
+ R \left(\frac{k}{k_{\mathrm{ref}}}\right)^{n_{2} - 1 
+ \frac{1}{2} \alpha_{2} \mathrm{ln} \left( \frac{k}{k_{\mathrm{ref}}} \right)
+\frac{1}{6} \beta_{2} \mathrm{ln}^2 \left( \frac{k}{k_{\mathrm{ref}}} \right)}
\right]~.
\label{eq:powerspectrum-two}
\end{eqnarray}
Here $A_s$ is the  total amplitude of the power spectrum and $R ( \equiv  {\cal P}_2 / {\cal P}_1$) 
is the amplitude of field 2 relative to that of the field 1. $n_i, \alpha_{i}$ and $\beta_i$ 
respectively are the spectral index, the running, and the running of the running for each field 
$(i = 1,2)$. These quantities are to be measured at the reference scale $k_{\rm ref}$ which we choose as $k_{\rm ref} = 0.05 \ {\mathrm{Mpc}}^{-1}$. The phenomenological form \eqref{eq:powerspectrum-two} captures models of two-field inflation as well as curvaton-type setups where the primordial perturbation may be sourced both by inflaton perturbations and perturbations of another component which during inflation was an isocurvature field.

We set $A_s$ equal to the observed best-fit amplitude of the curvature perturbation $A_{s} = \mathcal{P}_{\rm obs} = 2.19 \times 10^{-9}$~\cite{Ade:2015lrj} and let the other parameters vary in the range 
\begin{equation}  
\label{priors}
n_{1,2} \in [0, 2],  \  \alpha_{1,2} \in [-0.1, 0.1], \  \beta_{1,2} \in [-0.01, 0.01],  \ R \in [0, 1]~.
\end{equation}
We assume a flat prior distribution for all parameters. We choose the convention where ${\cal P}_1$ denotes the dominant part of ${\cal P}_{\zeta}$ and ${\cal P}_2$ the subdominant part such that $R\in[0,1]$ covers the full range of possible values. The priors chosen for $n_{1,2}$ allows for a significant scale-dependence $\Delta {\cal P}_{1,2}/{\cal P} \sim 10$ between the Planck pivot scale $k_{\rm ref}= 0.05~{\rm Mpc}^{-1}$ and the smallest scale probed by $\mu$ distortions  $k\sim 10^{4}~{\rm Mpc}^{-1}$~\cite{Khatri:2011aj} and most theoretical setups fall well within this range of $n_{1,2}$.  The priors for the running $\alpha_{1,2}$ and running of the running $\beta_{1,2}$ are chosen to be natural such that each successive term in the Taylor expansions of ${\cal P}_{1,2}$ is parametrically smaller than the previous one over the window $k = 0.05-10^4 ~{\rm Mpc}^{-1}$, as required by self-consistency of the expansion (\ref{eq:powerspectrum-two}). We have checked that our results do not essentially change if we double the prior range (\ref{priors}). This indicates that (\ref{priors}) represents a fair sample of the observationally allowed parameter space.

Using (\ref{eq:powerspectrum-two}), the spectral index, its running and the running of the running computed at a reference scale $k_{\rm ref}$ are given by
\begin{align}
n_{s} &\equiv  1 + \frac{\diff \mathrm{ln} \mathcal{P}_{\zeta}(k)}{\diff \mathrm{ln}(k)} \biggr\rvert_{k = k_{\mathrm{ref}}} = \frac{n_{1} + R n_{2}}{1 + R}, \label{eq:nseff_kref} \\
 \alpha_{s}&\equiv \frac{\diff n_{s}}{\diff \mathrm{ln}(k)} \biggr\rvert_{k = k_{\mathrm{ref}}} = \frac{\alpha_{1} + R \alpha_{2}}{1 + R} + \frac{R (n_{2} - n_{1})^{2}}{\left(1 + R\right)^{2}}, \label{eq:aseff_kref} \\
 \beta_{s}&\equiv \frac{\diff^{2} n_{s}}{\diff \mathrm{ln}(k)^{2}} \biggr\rvert_{k = k_{\mathrm{ref}}} = \frac{\beta_{1} + R \beta_{2}}{1 + R} +  \frac{3 R (n_{2} - n_{1}) (\alpha_{2} - \alpha_{1})}{(1 + R)^2} +  \frac{R (1 - R) (n_{2} - n_{1})^{3}}{(1 + R)^{3}} \label{eq:bseff_kref}~.
\end{align}
In particular, it should be noted that any difference between the individual spectral tilts of the two components, $n_1\neq n_2$, generates running and running of the running of the spectral index $n_{s}-1$. Moreover, configurations for 
which $n_1 = n_2$, $\alpha_{1} = \alpha_{2}$ and  $\beta_{1} = \beta_{2}$ are fully degenerate with the single field case $R=0$ for which $n_{s} = n_{1}, \alpha_{s}=\alpha_1$ and $\beta_{s} = \beta_1$.

%
\section{Model constraints from distortion measurements}\label{sec:singletwosignal}
%

We now move on to investigate the spectral distortion signals generated in the two-field case (\ref{eq:powerspectrum-two}). We scan over the seven model parameters $n_1,n_2, \alpha_1,\alpha_2, \beta_1,\beta_2, R$ in the prior range (\ref{priors}). For each parameter set we compute the spectral index $n_s$, its running $\alpha_s$, running of the running  $\beta_s$ and the spectral distortion $\mu$ using eqs. \eqref{eq:bseff_kref} and \eqref{eq:mu_integral}. We impose the Planck constraints~\cite{Ade:2015lrj}  \(n_{s} = 0.9586 \pm 0.0056\), \(\alpha_{s} = 0.009 \pm 0.010\) and \(\beta_{s} = 0.025 \pm 0.013\)   and confront the $\mu$ distortion against the forecasted 1-$\sigma$ sensitivity \(\Delta\mu = 1 \times 10^{-8}\) of the future PIXIE survey~\cite{pixie:2011}. Note that the presence of foregrounds can yield an order of magnitude degrade in PIXIE survey's sensitivity to \(\mu\)-distortion, as was discussed in \cite{abitbol:2017}.
  
\begin{figure}[tb!]
\begin{center}
\leavevmode
\includegraphics[scale=0.63]{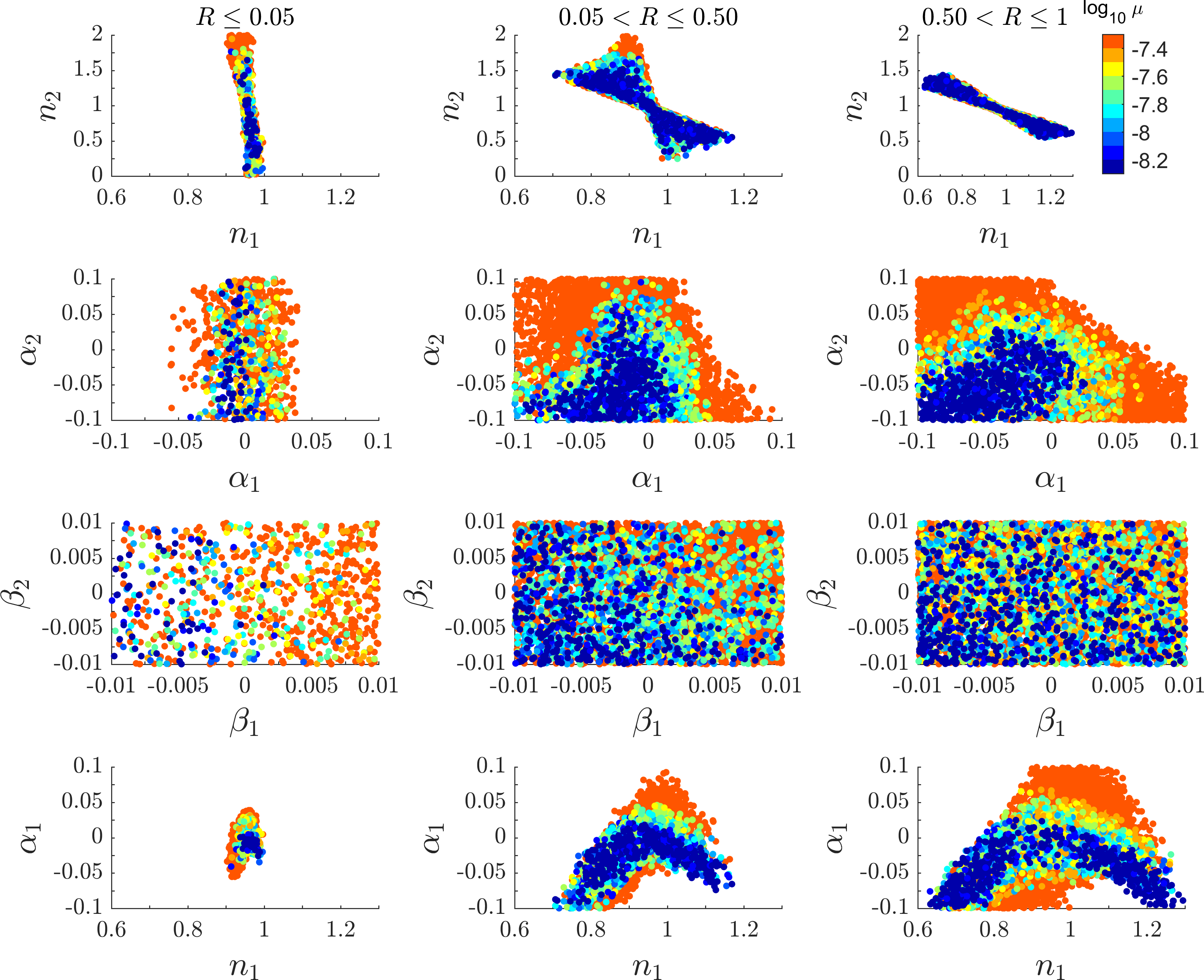}
\end{center}
\caption{Scatter plots showing the $\mu$ distortion (color coding) as function of the two-field model parameters compatible with the Planck bounds  \(n_{s} = 0.9586 \pm 0.0056\), \(\alpha_{s} = 0.009 \pm 0.010\) and \(\beta_{s} = 0.025 \pm 0.013\)  at 99 \% C.L. The limit $R=0$ corresponds to the single field case and for $R=1$ both fields contribute equally to the curvature perturbation. }
\label{fig:scatter_R_99}
\end{figure}

Our interest is to see how the combination of data can constrain the model parameters.
The results are shown in Figure~\ref{fig:scatter_R_99} which illustrates the dependence of $\mu$ on various combinations of  model parameters. Of the total scan of 5 000 000 parameter sets generated, 12 718 sets (0.25\%) were compatible with Planck bounds on (\(n_{s}, \alpha_{s}, \beta_{s}\)) at 99 \% C.L. We have also imposed the existing COBE/FIRAS bound on the spectral distortion $| \mu | < 9 \times 10^{-5}$~\cite{COBEFIRAS1,COBEFIRAS2}, which excludes only 90 of the sets compatible with Planck constraints. The $\mu$-values are positive apart from $9$ parameter sets for which the adiabatic cooling $\mu_{\rm ad} \approx -3 \times 10^{-9}$ generates a small negative $\mu$.  
%
%
%

Configurations with $R\leqslant 0.05$ shown in the first column of figure~\ref{fig:scatter_R_99} effectively reduce to the single field case. Apart from the enhanced $\mu$ values for $n_2\sim 2$, the subdominant field has no effect on the distortion signal in this class and the dependence of $\mu$ on $n_1,\alpha_1, \beta_1$ is that found in~\cite{Cabass:2016giw, Cabass:2016ldu}. The enhanced distortion signal is also induced from the positive running and running of the running due to the large difference $n_2 - n_1$ as seen from the last terms of eqs.~(\ref{eq:aseff_kref}) and (\ref{eq:bseff_kref}), which in this range are large enough to compensate for the $R$ suppression.
Note also that in this case $\mu$-distortion can be used to determine the sign of the running of running of the dominant field, $\beta_2$.

In the intermediate range, $0.05 < R \leqslant 0.5 $, we again see an enhancement of $\mu$ values for $n_2 \gtrsim 1.5$, but this feature vanishes for $R>0.5$ when both fields contribute roughly equally to the spectrum. Apart from this small enhancement, the distribution of $\mu$ values in the ($n_1, n_2$)-plane is uniform. On the other hand, the $\mu$ distortion depends strongly on individual runnings $\alpha_1$ and $\alpha_2$, whose allowed ranges will be considerably reduced by PIXIE sensitivity.  This can be traced back to the form of eqs.~(\ref{eq:aseff_kref}) and  (\ref{eq:bseff_kref}): when the difference between individual tilts $n_1$ and $n_2$ is large, the two terms in eq.~(\ref{eq:aseff_kref}) must cancel two keep the running $\alpha_s$ within the observational bounds. This induces a correlation between $n_i$ and $\alpha_i$, as illustrated for the dominant component on the last row of figure~\ref{fig:scatter_R_99}. Configurations with a large $(n_2-n_1)^2$ dependent part, which cancels against the $\alpha_{1,2}$ dependent part in (\ref{eq:aseff_kref}), in general lead to significant running of the running $\beta_s$ through the last two terms in (\ref{eq:bseff_kref}). This can easily dominate over the $\beta_{1,2}$ dependent first term and lead to a large positive or negative $\beta_s$ (and, respectively, large positive or negative $\mu$). This explains the flattening out of the distribution of $\mu$-values in the ($\beta_1,\beta_2$)-plane with increasing $R$, making $\mu$ less and less useful indicator of $\beta_i$.

\begin{figure}[tb!]
\begin{center}
\includegraphics[scale=0.38]{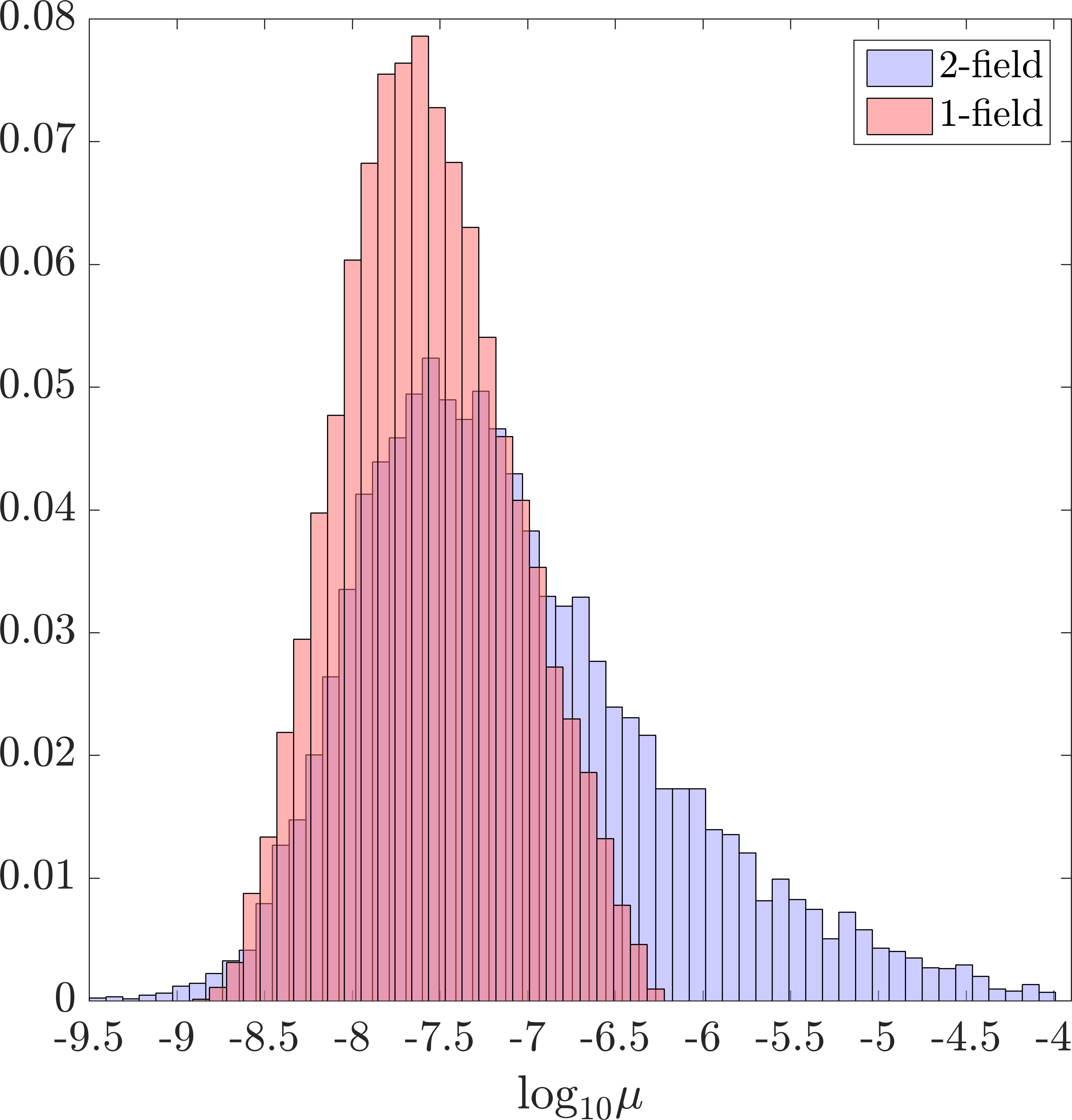} \qquad
\includegraphics[scale=0.38]{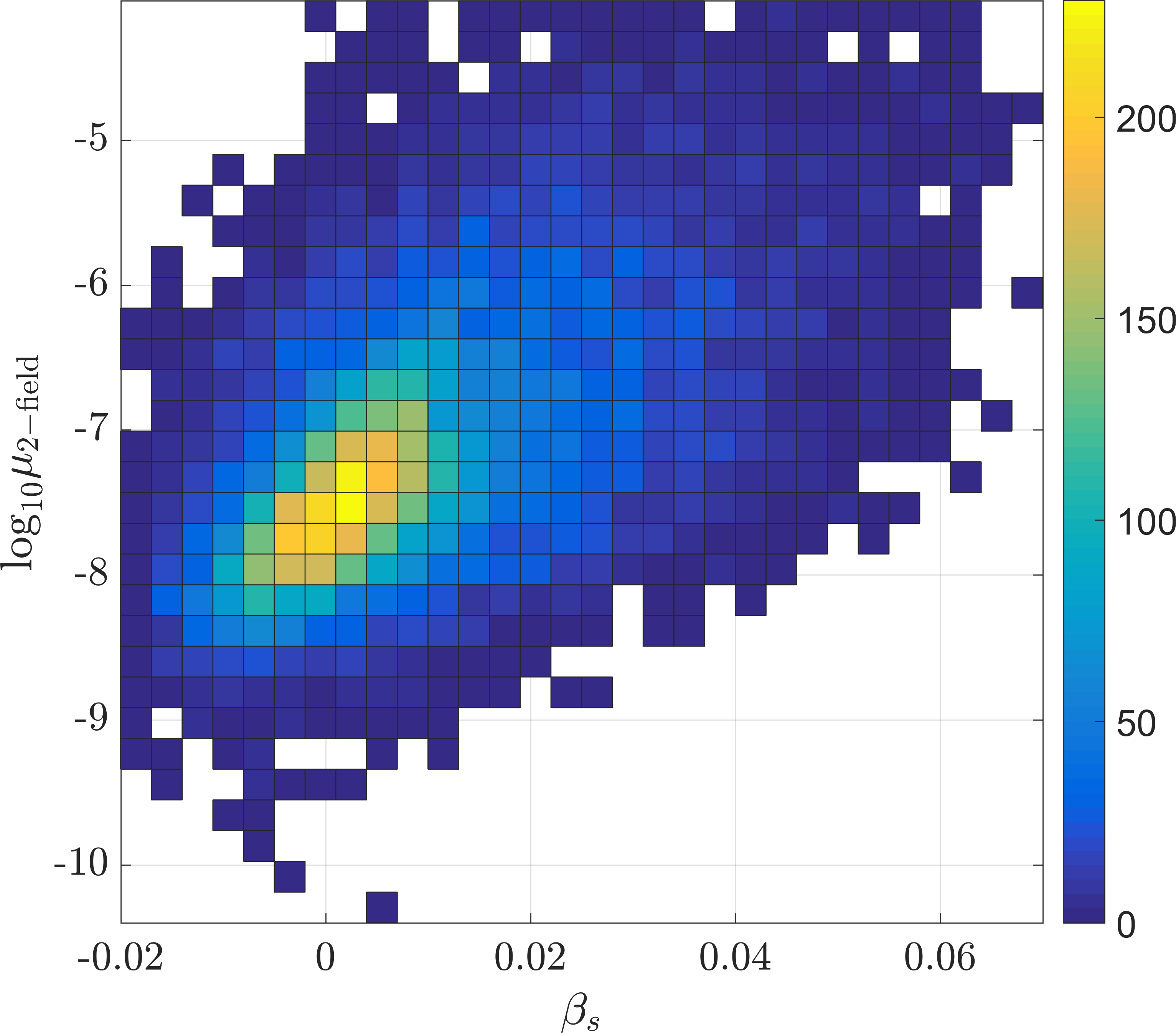}
\end{center}
\caption{Left panel: Normalized histograms of the $\mu$ distortion for configurations compatible with Planck constraints on (\(n_{s}, \alpha_{s}, \beta_{s}\)) at 99 \% C.L. and the COBE/FIRAS bound $| \mu | < 9 \times 10^{-5}$. The two cases shown correspond to the full general two-field case (blue) with the priors (\ref{priors}) and the single field limit $R=0$ with the same priors (red). Right panel: frequency distribution of two-field models corresponding to the blue histogram in the left panel as a function of $\beta_s$ and $\mu$.}
\label{fig:histogram_mus}
\end{figure}

The histogram in the left panel of fig.~\ref{fig:histogram_mus} shows the distribution of $\mu$-values for the accepted two field models in our scan, along with a similar distribution from a simulation in the single-field case ($R=0$), with the same prior ranges (\ref{priors}) for $n_1, \alpha_1, \beta_1$. The much wider spread of $\mu$-values in the two field case is due to the large running of the running induced by the $n_1,n_2, \alpha_1,\alpha_2$ dependent terms in eq.~(\ref{eq:bseff_kref}) discussed above. An observation of a large distortion, $\mu > 10^{-6}$, would be a compelling suggestion for a multifield inflation (unless our prior on single field $\beta$ is strongly underestimated). Recall also that we include only $\mu$ distortion from the Silk damping in our analysis and neglect any other possible energy injection processes which could also generate spectral distortions. In right panel we show the correlation of the running of running $\beta_s$ and the $\mu$ distortion for our accepted two-field models. The peak of the distribution, centered around $\beta_s=0$ is significantly shifted away from the Planck best fit range $\beta_{s} = 0.025 \pm 0.013$. This result is of course prior choice dependent and should not be given too strong a weight, albeit it is what follows with our natural choice of priors. One also sees that smaller (or negative) $\beta_s$ correlates with smaller $\mu$-values.
%
%
%

\begin{figure}[tb!]
\begin{center}
\leavevmode
\includegraphics[scale=0.63]{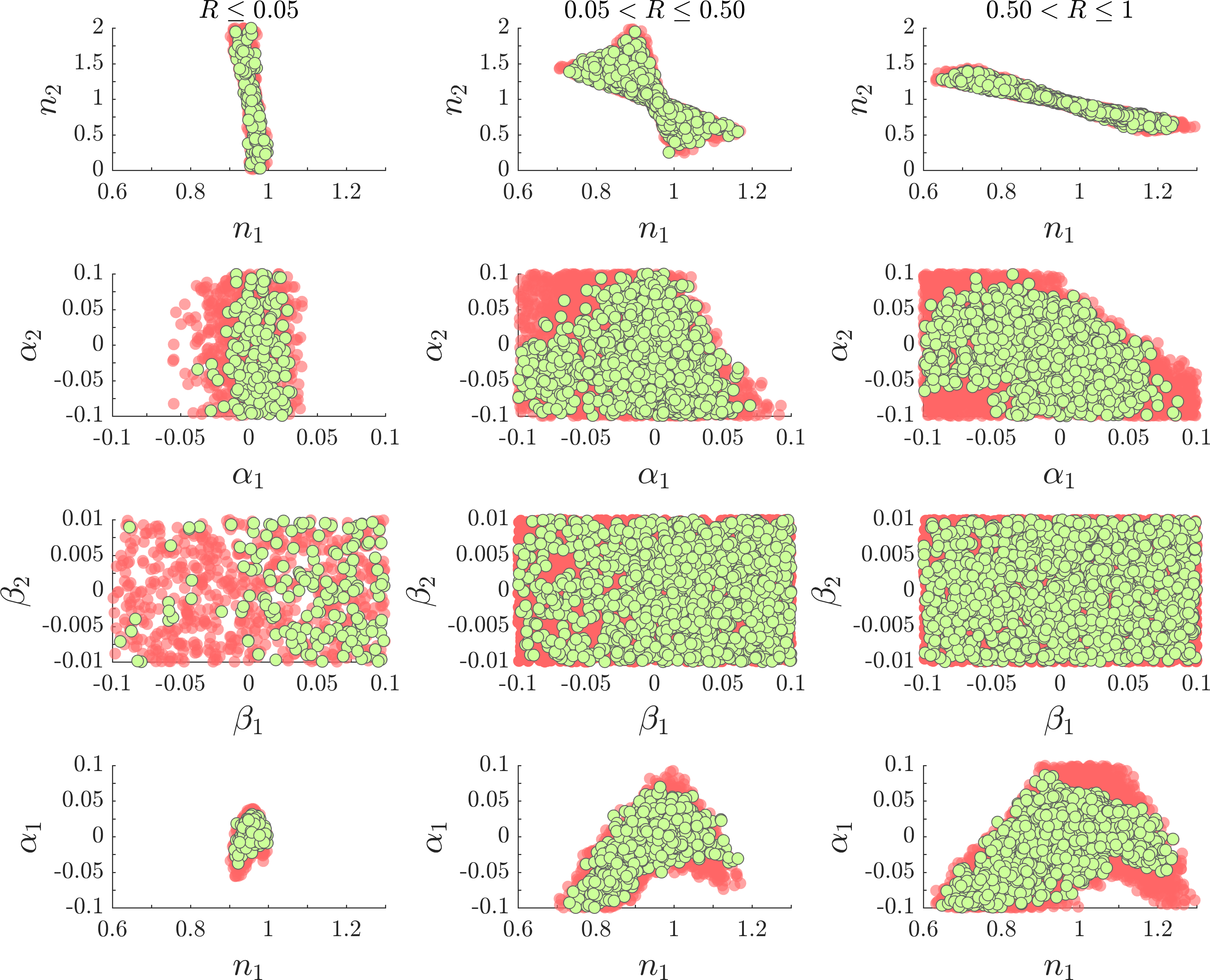}
\end{center}
\caption{Implications of an eventual detection of $\mu$ distortion for the two-field parameters. Red points show the entire parameter space compatible with Planck bounds on (\(n_{s}, \alpha_{s}, \beta_{s}, \mu\)) 99 \% C.L. Green points show the surviving 99 \% C.L. region assuming a detection \(\mu = (6 \pm 2) \times 10^{-8}\) at  95 \% C.L.}
\label{fig:scatter_R_msrmnt}
\end{figure}

Figure~\ref{fig:scatter_R_msrmnt} exemplifies consequences of an eventual detection assuming the measured spectral distortion would be \(\mu = (6 \pm 2) \times 10^{-8}\) at  95 \% C.L. Compared to figure~\ref{fig:scatter_R_99}, there are some differences. It can be seen that the detection would constrain $\alpha_1$ and $\alpha_2$ from above cutting away the configurations leading to large positive running and running of the running. Moreover, for  \(R \gtrsim 0.5\) also the tail of negative values \(\alpha_{1} \sim \alpha_{2} \sim -0.1\) gets cut out. For \(R \lesssim 0.05\), the most of the values of \(\beta_{1}\) are pushed to positive values \(\beta_{1} \gtrsim 0\). Overall,  the allowed parameter ranges do not change much however. Even a fairly clear detection of $\mu$ would then not help much to determine the individual power spectrum parameters in the general two field case. However, in specific theoretical setups the model parameters are typically more correlated, which should lead to a more precise determination of the allowed parameter ranges. We shall next show that this is indeed the case in a popular curvaton scenario.
%
%
%

%
\section{A specific example: mixed inflaton and curvaton scenario} \label{sec:modelcurv}
%

As a concrete example of a two-field model, we investigate the parameter space of the mixed inflaton curvaton scenario~\cite{Langlois:2004nn,Lazarides:2004we,Moroi:2005kz,Moroi:2005np,Ichikawa:2008iq,Fonseca:2012cj,Enqvist:2013paa,Vennin:2015vfa,Ichikawa:2008ne,Suyama:2010uj,Fujita:2014iaa}. Here the inflaton field $\phi$ is assumed to dominate the energy density during inflation. The curvaton $\chi$ is an energetically subdominant light iscourvature field during inflation but its fluctuations may later source curvature perturbation. In addition to the traditional incarnation of the curvaton scenario~\cite{Enqvist:2001zp,Lyth:2001nq,Moroi:2001ct} the relevant parts of our discussion here apply to any similar setup, such as modulated reheating~\cite{Dvali:2003em,Kofman:2003nx} or inhomogeneous end of inflation~\cite{Bernardeau:2004zz,Lyth:2005qk,Salem:2005nd} and so on, where a field which is energetically subdominant during inflation sources the adiabatic curvature perturbation well after the horizon exit of observable modes. 

We assume the scalar potential is of the form 
\beq
V(\phi,\chi) = V_1(\phi) + V_2(\chi)\ ,
\eeq 
and $V_1(\phi) \simeq 3H^2M_P^2 \gg V_2(\chi)$ during inflation, where $M_P$ is the reduced Planck energy scale. Furthermore, we assume that the non-negligible slow-roll parameters are given by
\baq
\epsilon_{\phi}  = \frac{M^{2}_P}{2}\frac{V^2_{,\phi}}{V^2},\;  
\eta_{\phi} = M^{2}_P \frac{V_{,\phi\phi}}{V},\;  
\eta_{\chi}= M^{2}_P \frac{V_{,\chi\chi}}{V},\; 
\xi_{\phi} = M^{4}_P \frac{V_{,\phi}V_{,\phi\phi\phi}}{V^2},\;
\sigma_{\phi} = M^{6}_P \frac{V_{,\phi}^2V_{,\phi\phi\phi\phi}}{V^3}. \;\;
\eaq
In other words, we assume that $\xi_{\chi} = M^{4}_{P} V_{,\chi} V_{,\chi\chi\chi}/V^2$ and $\sigma_{\chi} = M^{6}_{P} V_{,\chi}^2V_{,\chi\chi\chi\chi}/V^3$ are either identically zero, as is the case for the quadratic curvaton potential $V_{2} =m_{\chi}^2 \chi^2/2 $, or negligible due to the curvaton being close enough to isocurvature direction of the field space $V_{,\chi} \simeq 0$~\cite{Byrnes:2006fr}.

The spectrum of curvature perturbation is given by 
\beq
\label{Pzeta_curvaton}
{\cal P}_{\zeta}(k) = N_{,\phi}^2(k) {\cal P}_{\phi}(k) +  N_{,\chi}^2(k) {\cal P}_{\chi}(k)
\eeq
where the derivatives of the number of $e$-folds can be evaluated at the horizon crossing of the mode $k = aH$. To leading order in slow-roll we have  
\beq
\label{Pphichi_Nphi}
{\cal P}_{\phi}(k)={\cal P}_{\chi}(k) = \left(\frac{H}{2\pi}\right)_{k = aH}^2~, 
\qquad N_{,\phi}(k) = \frac{1}{M_{P} \sqrt{2\epsilon(k)}}~,
\eeq
whereas $N_{,\chi}$ depends on details of the curvaton setup. Defining 
\beq
R = \frac{N_{,\chi}^2(k_{\rm ref})}{N_{,\phi}^2(k_{\rm ref})}~, 
\eeq
the spectrum  (\ref{Pzeta_curvaton}) coincides with the phenomenological two-field form (\ref{eq:powerspectrum-two}). Note a notational difference however, here $R\in [0,\infty]$ as the curvaton can give either a subdominant or dominant contribution to the total spectrum whereas in (\ref{eq:powerspectrum-two}) we defined $R$ to be the ratio of the subdominant and dominant components of ${\cal P}_{\zeta}$. Equation (\ref{Pzeta_curvaton}) at the reference scale $k_{\rm ref}$ can then be recast in the form 
\beq
\label{Pzeta_PT}
{\cal P}_{\zeta}(k_{\rm ref}) = {\cal P}_{T}(k_{\rm ref}) \frac{1+ R}{16 \epsilon(k_{\rm ref})}~,
\eeq
where ${\cal P}_T = 8 H^2/(4 \pi^{2} M_P^2)$ is the spectrum of gravitational waves. The tensor-to-scalar ratio at $k_{\rm ref}$ given by 
\beq
r \equiv \frac{{\cal P}_T}{{\cal P}_{\zeta}}=\frac{16 \epsilon}{1+R}~.
\label{eq:rtensor}
\eeq

The spectral index, its running and running of the running are given by eqs. (\ref{eq:nseff_kref}), (\ref{eq:aseff_kref}) and (\ref{eq:bseff_kref}) where $n_1 = n_{\phi},~n_2 = n_{\chi}$ etc. and the explicit expressions of these parameters are given by 
\begin{align}
n_{\phi} &= 1 - 6 \epsilon_{\phi} + 2 \eta_{\phi}~, \label{eq:SRspectral} \\
\nonumber
n_{\chi} &= 1 - 2\epsilon_{\phi}+2\eta_{\chi}~,  \\
\nonumber
\alpha_{\phi} &= -24 \epsilon_{\phi}^2 + 16 \epsilon_{\phi} \eta_{\phi} - 2 \xi_{\phi}~, \\
\nonumber
\alpha_{\chi} &= -8\epsilon_{\phi}^2+4\epsilon_{\phi}\eta_{\phi}+4\epsilon_{\phi}\eta_{\chi}~, \\
\nonumber
\beta_{\phi} &=  2\sigma_{\phi} -8 \epsilon_{\phi}(4\eta_{\phi}^2+3\xi_{\phi})+2\eta_{\phi}\xi_{\phi} +192\epsilon_{\phi}^2\eta_{\phi}-192\epsilon_{\phi}^3~,\\
\nonumber
\beta_{\chi} &=-64 \epsilon_{\phi}^3 - 8\epsilon_{\phi}\eta_{\phi}(\eta_{\phi}+\eta_{\chi})  +8 \epsilon_{\phi}^2( 7\eta_{\phi}+ 3\eta_{\chi})
- 4 \epsilon_{\phi} \xi_{\phi}~.
\end{align}

\subsection{Scan over model parameters}
%

We fix the spectrum at the observed value ${\cal P}_{\zeta} = 2.19 \times 10^{-9}$~\cite{Ade:2015lrj} at the pivot scale and impose the Planck+BICEP2/KECK bound $r < 0.07$ (at 95~\% CL)~\cite{Array:2015xqh}. We scan over the slow roll parameters $\epsilon,\eta_{\phi}, \eta_{\chi}, \xi_{\phi},\sigma_{\phi}$ and $R$ with the prior ranges  
\begin{eqnarray}
&&\epsilon_{\phi} \in [0,0.1], \qquad
\eta_{\phi}, \eta_{\chi} \in [-0.1,0.1], \qquad
\xi_{\phi} \in [-0.01,0.01], 
\nonumber\\ 
&& \sigma_{\phi} \in [-0.001,0.001], \qquad
R \in [0.01,100]~. 
\label{eq:SRpriors1}
\end{eqnarray}
The prior range for $R$ practically extends from the inflaton dominated limit to curvaton domination and increasing the range further would not affect our results. On the other hand, the priors chosen for the slow-roll parameters do impose some constraints on the models included. We chose to concentrate on models where both the curvaton and the inflaton dynamics is described by the slow roll dynamics, but allowed for relative large values for the slow-roll parameters, up to $\epsilon, |\eta| \sim 0.1$. Of the higher order slow-roll parameters $\xi$ and $\sigma$ we assumed that they are at most of order $\epsilon^2$ and $\epsilon^3$, respectively. Note that for each parameter choice the inflationary scale $H$ is fixed by eq.~(\ref{Pzeta_PT}) as we fix ${\cal P}_{\zeta}$ to the observed amplitude. 

%
%
\begin{figure}[tb]
\begin{center}
\leavevmode
\includegraphics[scale=0.55]{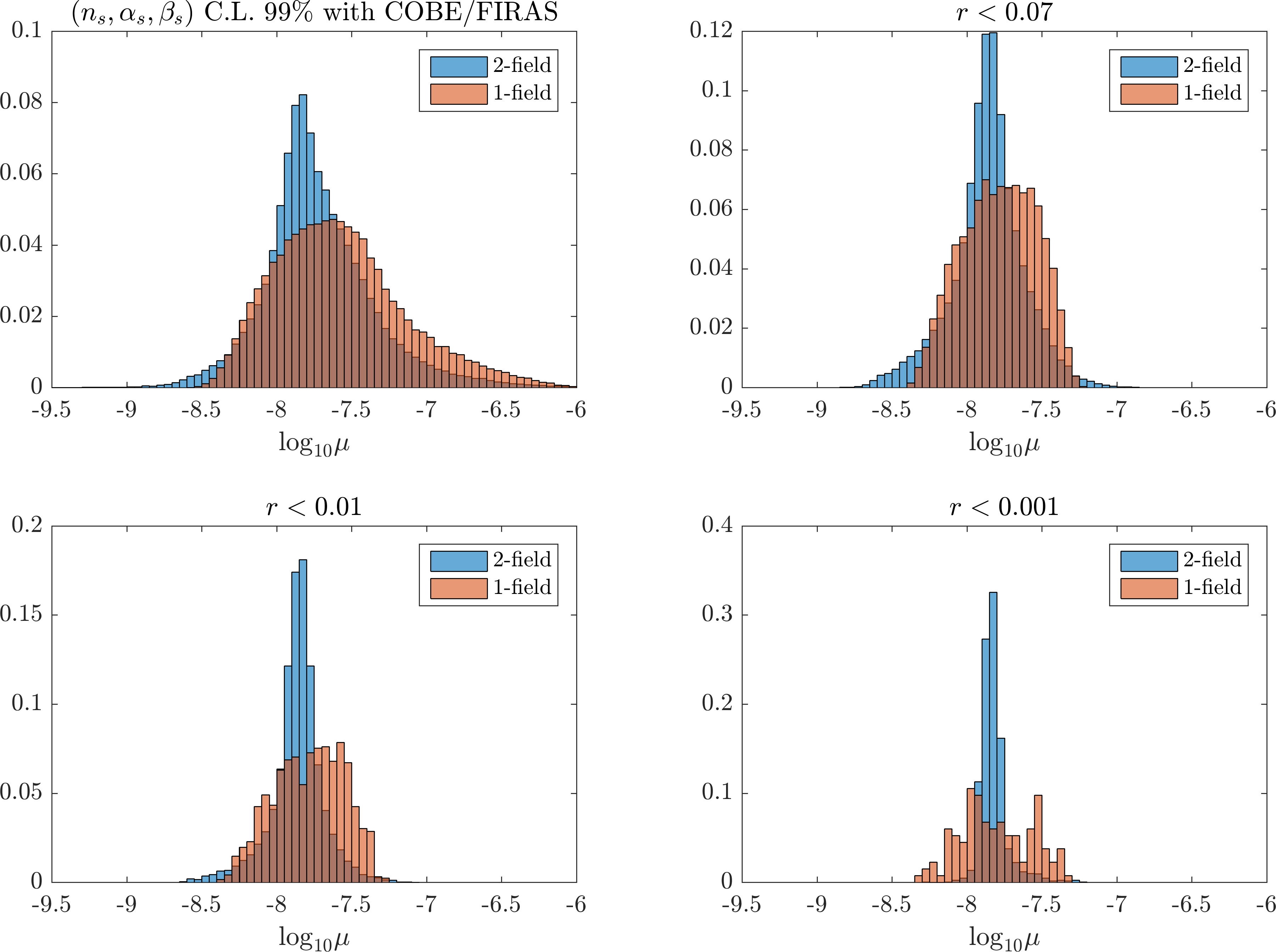}
\end{center}
\caption{Histograms showing the distribution of $\mu$ values for the mixed inflaton-curvaton model (blue) with the priors \eqref{eq:SRpriors1} and for single-field inflation (red) $R=0$ with the same priors. All configurations shown are compatible with Planck bounds on (\(n_{s}, \alpha_{s}, \beta_{s}\)) at 99 \% C.L. } 
\label{fig:SRcurv_histogramtensor}
\end{figure}
%

We then impose Planck bounds on (\(n_{s},\alpha_{s},\beta_{s}\)), compute the distortion signature for the accepted models and impose also the COBE/FIRAS $\mu$-bound. Of the 3 000 000 models created in our scan, 121 086 models (4\%) survived the bounds. The distributions of $\mu$-values in the accepted set is shown in the histogram~\ref{fig:SRcurv_histogramtensor}, along with results of an analogous single field scan. First panel shows distributions with no bound on the tensor-to-scalar ratio $r$, while others show the effect of imposing varying priors on $r$. This is crucial because a constraint on $r$ imposes an $R$-dependent upper bound on $\epsilon_{\phi}$ through eq. (\ref{eq:rtensor}). This removes configurations where large positive $\eta_{\phi}$ or $\eta_{\chi}$ cancel in eq. (\ref{eq:SRspectral}) against large negative terms proportional to $\epsilon_{\phi}$ to yield the observed spectral index. Second panel shows that already the existing bound $r < 0.07$~\cite{Array:2015xqh} (49 632 models survive) cuts out large distortions. The cut is most severe in the pure inflaton limit $R=0$ where $n_s = 1 -6 \epsilon_{\phi} + 2 \eta_{\phi}$ and setting $r = 16\epsilon_{\phi} < 0.07 $ removes most of positive $\eta_{\phi}$ allowed by the priors. This cuts out large $\mu$ values as positive $\eta_{\phi}$ contribute positively to the runnings $\alpha_{\phi}$ and $\beta_{\phi}$ in eq. (\ref{eq:SRspectral}) and hence generate large distortions. In the opposite limit of  curvaton domination, $R \gg 1$, the tilt feels only $\eta_{\chi}$ but the runnings and $\mu$ distortion are affected also by $\eta_{\phi}$ which is now essentially unconstrained. The intermediate case where both the inflaton and curvaton contributions are relevant is a mixture of these limits. As a result, we find that after imposing the bound on $r$ the range of possible $\mu$ values in the mixed inflaton curvaton case extends to both smaller and also to slightly larger values compared to the pure inflaton case with the same slow roll priors. This is also seen in fig. \ref{fig:curvaton_powerspectrum} below, which shows the range of possible power spectrum shapes ${\cal P}_{\zeta}(k)$ in both cases. The presence of two sources in the mixed case allows for larger runnings which results the greater spread of $\mu$ values. Interestingly, imposing more stringent priors on $r$ does not help constraining the $\mu$ range more\footnote{Because one-field model is a subset of the two-field model its distribution must always be narrower than the two-field distribution. The latter is significantly more peaked however, with relatively few points at the tails. The very narrow two-field distribution in case \(r < 0.001\) is thus but an artefact due to the smallness of the surviving sample.}. However, more information can be obtained when $r$ prior is applied to $\mu$ correlated with pairs of model parameters. 

%
%
\begin{figure}[tb]
\begin{center}
\leavevmode
\includegraphics[scale=0.55]{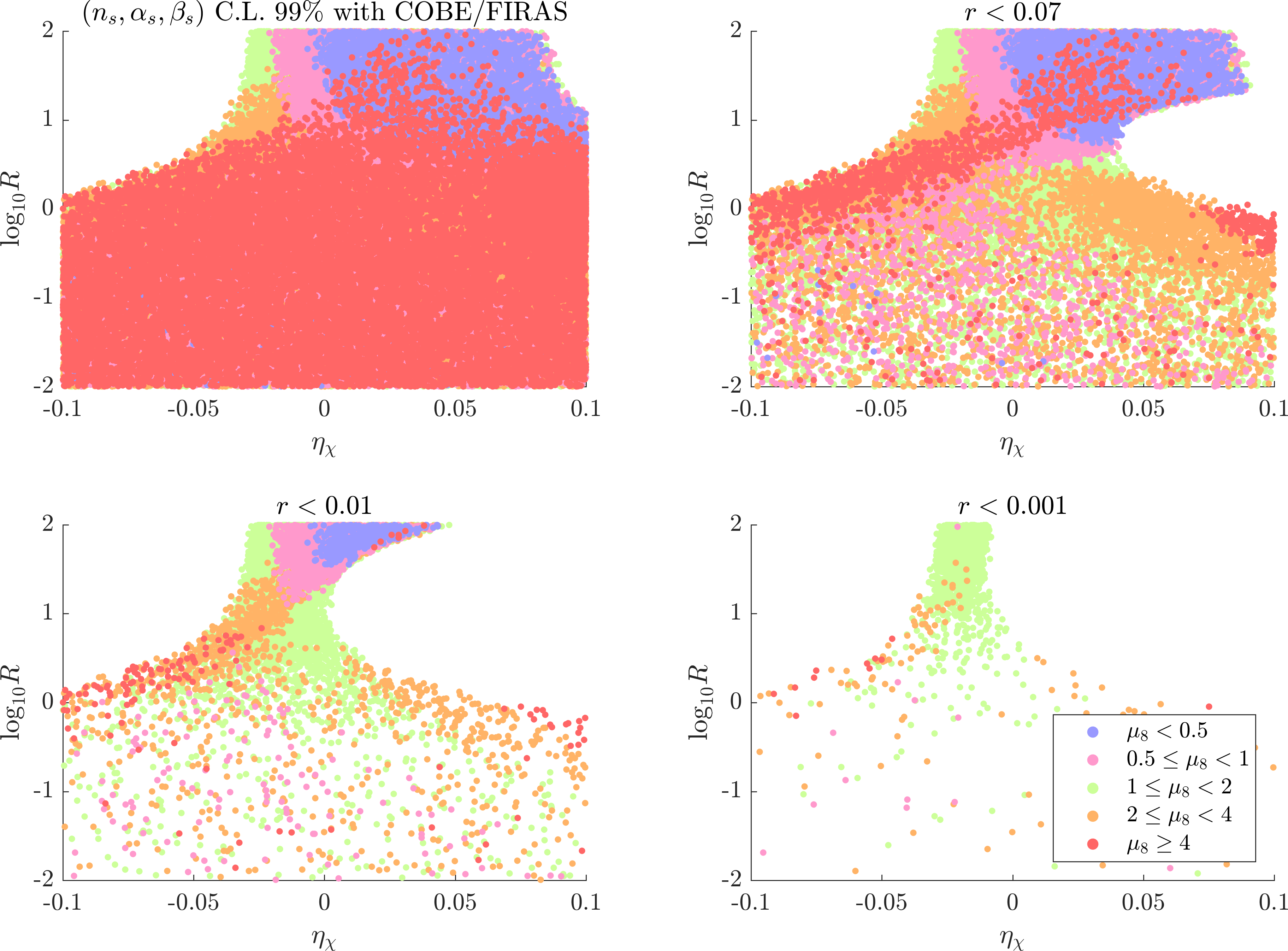}
\end{center}
\caption{
Scatter plots showing the $\mu$ distortion (color coding) in the mixed inflaton-curvaton setup as function of the curvaton parameters $\eta_{\chi}$ and $R$. The depicted points are compatible with the Planck bounds~\cite{Ade:2015lrj}  \(n_{s} = 0.9586 \pm 0.0056\), \(\alpha_{s} = 0.009 \pm 0.010\) and \(\beta_{s} = 0.025 \pm 0.013\)  at 99 \% C.L. The limit $R=0$ corresponds to inflaton domination and $R\gg 1 $ curvaton domination. Here $\mu_{8} = \mu \times 10^{8}$.}
\label{fig:SRcurv_REtachi_tensor}
\end{figure}
%

Figure~\ref{fig:SRcurv_REtachi_tensor} shows the dependence of $\mu$ on the curvaton parameters \(\eta_{\chi}\) and \(R\). The first panel again depicts the results without the $r$-bound. In this case the $\mu$ values are strongly degenerate on the \((\eta_{\chi}, R)\) plane. The corners of large negative and positive \(\eta_{\chi}\) for large \(R\) are excluded as they produce too red or blue spectrum respectively.  Taking the current tensor bound $r < 0.07$ into account (second panel) cuts out large $\mu$ values in the inflaton dominated regime $R\lesssim 1$, as explained above. Configurations around $R\sim 1$ and $\eta_{\chi}\gtrsim 0.05$ get removed as obtaining the observed spectral index in this regime would require $\eta_{\phi} < -0.1$ for $r < 0.07$, which is outside our prior range. Decreasing the tensor-to-scalar ratio further increases the size of this cut-out region, as is clearly seen in the two lower panels. The $\mu$ distortion is not completely fixed by $\eta_{\chi}$ and $R$ even in the curvaton dominated limit $R \gg 1 $ but depends also on the inflaton slow-roll parameters (which measure time dependence of the Hubble rate). Even with this degeneracy left, the results indicate that a measurement of spectral distortions together with tensor-to-scalar ratio could place quite non-trivial constraints on the curvaton parameters\footnote{We have checked that for $\mu \lesssim  4 \times 10^{-8}$ the results are robust against increasing the prior ranges. The distribution of larger $\mu$-values on the other hand does change if the priors are extended to allow for significant deviations from slow-roll, $\epsilon_{\phi}, |\eta_{\phi}|\gtrsim 0.2$.}. These directly translate into constraints on the shape of the power spectrum ${\cal P}_{\zeta}(k)$ as illustrated in the right panel of fig. \ref{fig:curvaton_powerspectrum}. As can be seen in Figure \ref{fig:SRcurv_REtachi_tensor} the degeneracy between $\mu$ and the curvaton parameters $\eta_{\chi}$ and $R$ decreases the tighter the bound on $r$ becomes. This is because for the priors (\ref{eq:SRpriors1}) the degeneracy is mostly due to $\eta_{\phi}$ terms in the runnings (\ref{eq:SRspectral}) and they are all multiplied by $\epsilon_{\phi}$.

%
%
\begin{figure}[tb!]
\begin{center}
\includegraphics[scale=0.38]{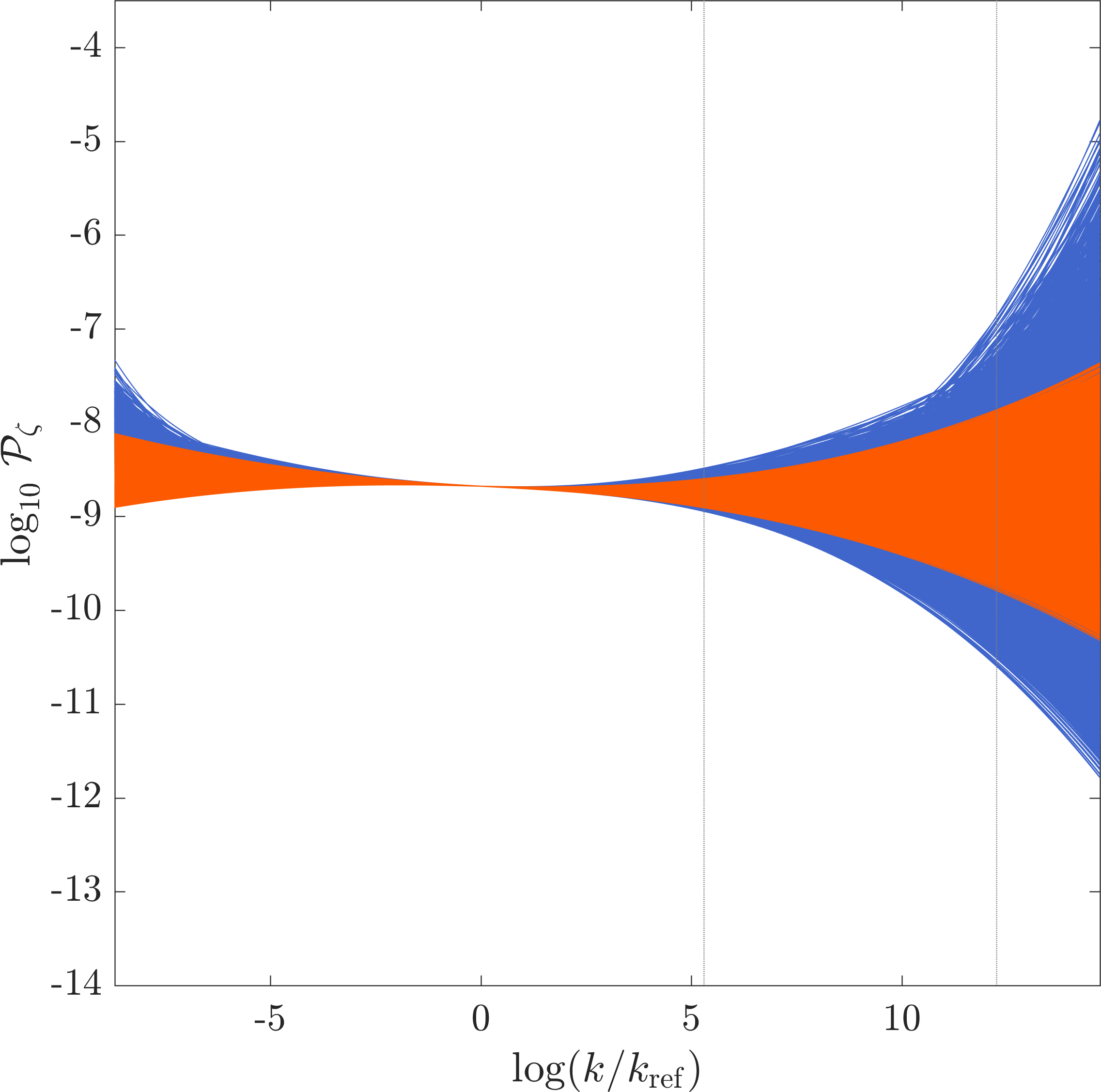} \qquad
\includegraphics[scale=0.38]{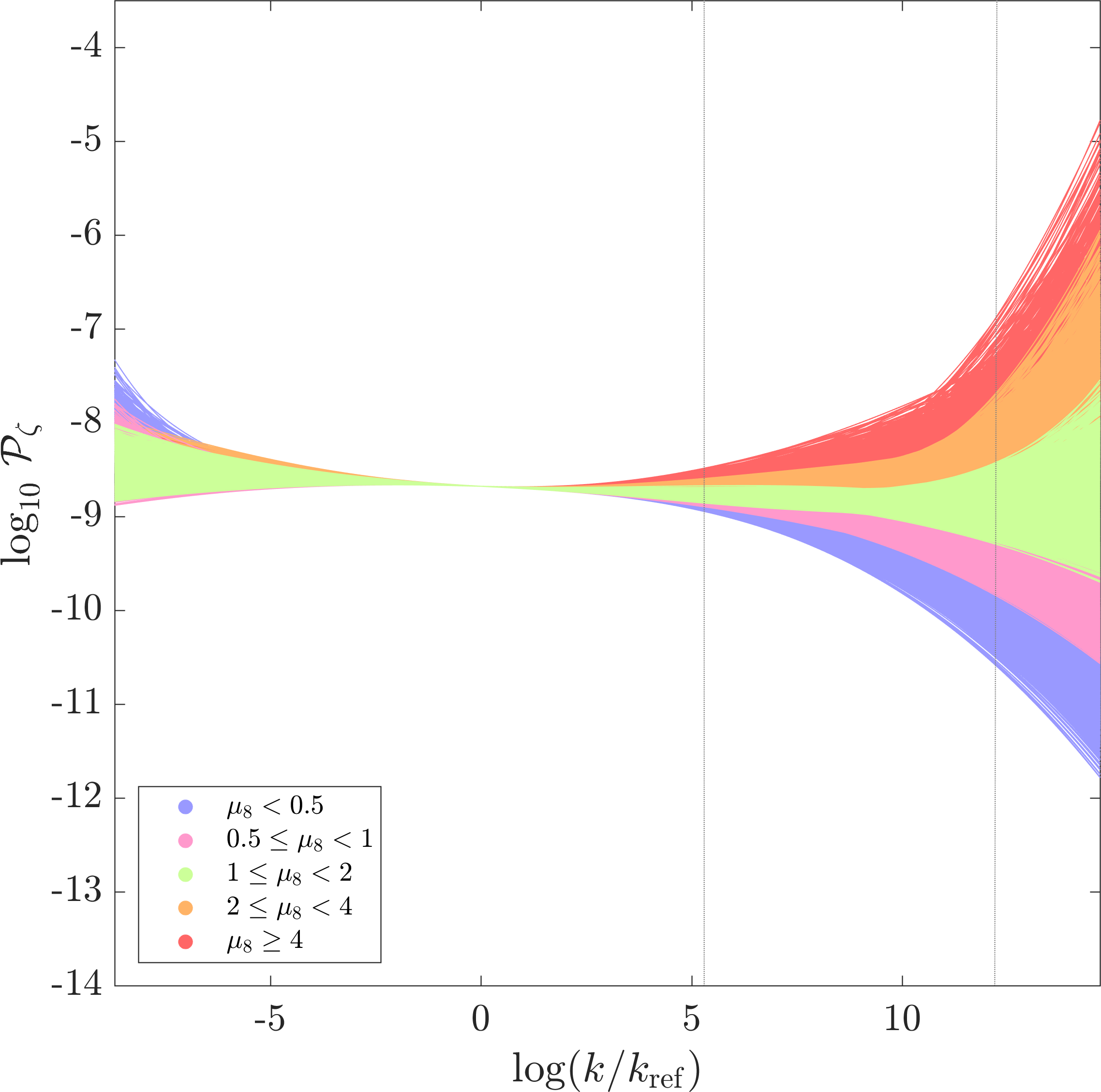}
\end{center}
\caption{The left panel shows the spread of possible power spectrum shapes $\mathcal{P}_{\zeta}(k)$ in the mixed inflaton-curvaton case (blue) and in the pure inflaton case  (orange). The right panel shows $\mu$-distortion values in the mixed case corresponding to different shapes of the spectrum. The shown configurations are compatible with Planck constraints on $(n_{s},\alpha_{s},\beta_{s})$ at 99 \% C.L. and the tensor bound $r < 0.07$. The dashed lines at $k \sim 10 \ \mathrm{Mpc}^{-1}$ and $k \sim 10^{4} \ \mathrm{Mpc}^{-1}$ represent the range which the spectral \(\mu\)-distortion is most sensitive to.}
\label{fig:curvaton_powerspectrum}
\end{figure}
%

In summary, these results show that distortion constraints are especially powerful when combined with tighter bounds on tensor-to-scalar ratio, alredy at the level of the forecasted sensitivity of the next generation polarization probes (see, e.g.,~\cite{Barron:2017kuo}). It would be interesting to see if other observables, such as non-Gaussianity would be similarly useful when combined with distortion. For example, in~\cite{Byrnes:2014xua}, observational constraints on the simplest inflaton curvaton model were investigated using the spectrum of curvature perturbations and  non-Gaussianity. It would be interesting to redo a similar analysis including spectral distortions in the study.

\section{Conclusion} 
\label{sec:conclusion}
%

We have investigated the CMB spectral $\mu$ distortion in generic two-field models 
where two uncorrelated sources can simultaneously contribute to the primordial power spectrum.
This kind of models can be characterised by the total amplitude at the pivot scale $A_s$, the ratio 
$R$ $(\equiv  {\cal P}_2 / {\cal P}_1)$ of the amplitude between two fields, the spectral index 
$n_i$, the running $\alpha_i$ and the running of the running $\beta_i$ for each field $(i=1,2)$.
We scanned the model parameter space with broad priors and computed the $\mu$ distortion signal for
parameters consistent with the Planck observations. We found that the value of the $\mu$ 
distortion are enhanced when (i)~$n_2$ is large (we have adopted the prior range $0 \le R \le 1$), 
(ii)~the difference between $n_2$ and $n_1$ or $\alpha_2$ and $\alpha_1$ is large, while satisfying 
the Planck constraints on $n_s, \alpha_s$ and $\beta_s$, as seen in Fig.~\ref{fig:scatter_R_99}. 
These two cases arise because of the two-field nature of the model:  even if the field 2 is 
subdominant at CMB scales, it can give a dominant contribution, especially when $n_2$ is blue-tilted
on small scales where the $\mu$ distortion can probe the primordial fluctuations, which was also 
discussed in~\cite{Enqvist:2015njy}. The second point~(ii) is due to the fact that large differences $n_2 - n_1$ \cite{kinney_2012} and/or $\alpha_2 - \alpha_1$ can 
induce large positive running and  running of the running as read off from 
eqs.~\eqref{eq:aseff_kref} and \eqref{eq:bseff_kref}. We have also made an analysis where the  
detection of $\mu$ distortion is assumed in a future observations such as PIXIE~\cite{pixie:2011} 
(or something analogous to PRISM~\cite{PRISM1,PRISM2}), say \(\mu = (6 \pm 2) \times 10^{-8}\) at  
95 \% C.L., then we found that  we could exclude some models, especially where the difference 
between $\alpha_1$ and $\alpha_2$ are large,  even if the constraints on $n_s, \alpha_s$ and 
$\beta_s$ from CMB  are satisfied, which is shown in Fig.~\ref{fig:scatter_R_msrmnt}.

As concrete example of a two-field model, we studied the mixed inflaton-curvaton scenario and found that the $\mu$ distortion can efficiently break degeneracies among the curvaton parameters. For a very small $\mu$, the distortion signal in this scenario is tightly correlated with the curvaton slow-roll parameter $\eta_{\chi}$ and the curvaton contribution to primordial perturbations $R$, and the correlation becomes stronger the tighter the upper bound on the tensor-to-scalar ratio $r$. This indicates that future observations or improved bounds on the $\mu$ distortion and primordial gravitational waves could imply a severe test for this kind of multi-field models. We expect that correlating $\mu$ distortion similarly with various other observables, such as non-Gaussianities, would give further useful tests for inflationary models.

\acknowledgments
TT would like to thank  Cosmology group at University of Jyv\"{a}skyl\"{a} for the hospitality during the visit, where a part of this work was done. This work was partially supported by the Jenny and Antti Wihuri foundation (JL), the Academy of Finland project 278722 (KK), JSPS KAKENHI Grant Number 15K05084 (TT), and MEXT KAKENHI Grant Number 15H05888 (TT).

\end{document}